# Magnetism and effect of anisotropy with one dimensional monatomic chain of cobalt by a Monte Carlo simulation

Lin He, Desheng Kong and Chinping Chen[*]

Department of Physics, Peking University, Beijing, 100871, People's Republic of China

ABSTRACT

   The magnetic properties of the one dimensional (1D) monatomic chain of Co reported in a previous experimental work are investigated by a classical Monte Carlo simulation based on the anisotropic Heisenberg model. In our simulation, the effect of the on-site uniaxial anisotropy, $K_u$, on each individual Co atom and the nearest neighbour exchange interaction, $J$, are accounted for. The normalized coercivity $H_C(T)/H_C(T_{CL})$ is found to show a universal behaviour, $H_C(T)/H_C(T_{CL}) = h_0 ( e^{T_B/T} - e )$ in the temperature interval, $T_{CL} < T \leq T_B^{Cal}$, arising from the thermal activation effect. In the above expression, $h_0$ is a constant, $T_B^{Cal}$ is the blocking temperature determined by the calculation, and $T_{CL}$ is the temperature above which the classical Monte Carlo simulation gives a good description on the investigated system. The present simulation has reproduced the experimental features, including the temperature dependent coercivity, $H_C(T)$, and the angular dependence of the remanent magnetization, $M_R(\theta,\phi)$, upon the relative orientation ($\theta,\phi$) of the applied field $H$. In addition, the calculation reveals that the ferromagnetic-like open hysteresis loop is a result of a slow dynamical process at $T < T_B^{Cal}$. The dependence of the dynamical $T_B^{Cal}$ on the field sweeping rate $R$, the on-site anisotropy constant $K_u$, and the number of atoms in the atomic chain, $N$, has been investigated in detail.

## 1. Introduction

One dimensional (1D) spin lattice models have been studied in statistical physics for more than 80 years owing to a theoretical interest following the pioneer work of Ising [1]. The research interest of such systems lies in the physical properties, such as the phase transition, the dynamic process, *etc.*, showing distinctive features from the bulk material. Relevant investigations have their significances not only for the phenomena with the 1D system but also for a better understanding of the properties of three dimensional (3D) bulk material [2,3]. One of the most important results obtained with these theoretical investigations is the Mermin-Wagner theorem stating that a long-range order of an infinite linear chain is absent at a finite temperature with a short-range exchange interaction (SREI) [4-5].

Traditionally, a truly 1D system is untenable in experiment and thus bulk materials containing chain-like structure with strong magnetic intrachain interaction, $J$, along one direction are considered as a realizable 1D system [2]. Such a strong 1D magnetic intrachain coupling is usually produced by separating the chain-like structure with large non-magnetic ions or complexes. The interchain interaction, $J'$, is therefore much weaker than the intrachain interaction, $J$, with these materials. Consequently, the properties arising from the 1D effect become significant only within the temperature interval $J'/k_B < T < J/k_B$, where $k_B$ is the Boltzmann constant, and at $T \leq J'/k_B$, the 3D properties take over gradually [2]. For a 1D magnetic system realized with such a bulk material, it is usually considered as an infinite 1D chain. Long range magnetic ordering is therefore not expected with such a system according to the Mermin-Wagner theorem. However, it is also possible to observe ferromagnetic (FM) like properties, for example, an open hysteresis loop, in a slow dynamical process with these magnetic 1D systems [6] arising from the presence of a large uniaxial anisotropy. The possibility of generating 1D magnetic chains with long relaxation time at room temperature is very valuable for industrial applications and so relevant investigations have become an active research field [6].

An alternative approach to generate a 1D magnetic system has been made possible by the rapid advancements in material synthesis technology over the past few decades to artificially construct 1D chains on a nonmagnetic substrate [7-9]. Progress has been brought out by the work of Gambardella, *et al.*, reported recently [10]. They have fabricated a 1D Co chain with a length of 80 atoms arranged in parallel on a Pt (997) step edge. Two major magnetic properties have been observed experimentally, a) the existence of a FM like open hysteresis loop below the blocking temperature, $T_B^{Exp}$ = 15 ± 5 K, and b) the angular dependence of the remanent magnetization, $M_R(\theta,\phi)$, upon the orientation of the applied field, $H$. This experimental result is unexpected by treating the 1D chain of 80 atoms as an infinite chain with a SREI in equilibrium, according to the Mermin-Wagner theorem [4,5]. It is noted that the Mermin-Wagner theorem actually does not apply to a finite spin chain due to the lack of translational symmetry. Recently, Denisov and Hänggi [11] have studied a finite Ising spin chain with a nearest neighbour exchange interaction. They found that there exists a characteristic temperature below which a finite spin chain would exhibit an FM behaviour if the Ising spins are treated as an approximation of the classical

Heisenberg spins with the presence of a very large uniaxial anisotropy potential between the two equilibrium states. Nevertheless, the natural characteristics of the open hysteresis loop as well as the remanent magnetization observed in the experiment are nonequilibrium phenomena involving slow relaxation process [10]. Recently, Li and Liu have applied a kinetic Monte Carlo (KMC) method [12,13] to investigate this 1D monatomic Co chain system. They have explicitly started with an anisotropic Heisenberg model and reduced it analytically to an Ising one subject to a potential barrier at the presence of an uniaxial anisotropy. The transition state of the spin flips between two metastable states owing to the presence of a potential barrier was accounted for. The numerical calculation by the KMC method has then been performed. They have reproduced the FM feature of an open hysteresis loop at $T = 10$ K, and a superparamagnetic (SPM) behaviour at $T = 45$ K. According to the above analysis [11,12], the presence of a potential barrier resulting from a large uniaxial anisotropy is essential to the existence of FM properties for the 1D monatomic chain of Co with a SREI.

In order to further understand the effect of anisotropy on the magnetization reversal of the 1D Co chain system in the experiment performed by Gambardella *et.al* [10,14-16], we have applied the classical MC method on the anisotropic Heisenberg model without resorting to the approximation to the Ising model at the presence of a large potential barrier as carried out by Li, *et al.* [12]. The MC simulation adopts the standard Metropolis algorithm with a random updating scheme, see for example [17,18]. This method has been applied to calculate the nonequilibrium dynamical phase transition properties of an anisotropic Heisenberg ferromagnet driven by an elliptically polarized magnetic field with success [19]. In our simulation, the numerical results have reproduced the magnetic behaviours of the 1D monatomic chain of Co observed in the experiment, including the *M-H* behaviour at $T = 10$ and 45 K and the angular dependence of the remanent magnetization, $M_R(\theta,\phi)$ upon the relative orientation of the applied field. Further calculations have revealed some interesting properties beyond the results obtained in the experiments [10] and in the calculations [12]. In particular, $H_C(T)$ measured along the axis of anisotropy can be described by the function $H_C(T)/H_C(T_{CL}) = h_0 ( e^{T_B/T} - e)$ within the temperature range, $T_{CL} < T \leq T_B^{Cal}$, where $T_B^{Cal}(N,R,K_u)$ is the blocking temperature depending on the number of atoms in the atomic chain, $N$, the field sweeping rate, $R$, and the anisotropy constant, $K_u$, and $T_{CL}$ is the temperature limit above which the classical MC method describe the property of thermal activation properly. The dependence of $T_B^{Cal}$ on $N$ with a fixed $K_u$ has been calculated also. The result explicitly demonstrates that $T_B^{Cal}$ depends on the chain length with a small $N$ and it approaches a constant value with $N$ exceeding a critical value $N_C$.

## 2. Calculation procedure

The classical Heisenberg model with an on-site uniaxial anisotropy perpendicular to the chain axis was adopted to calculate the magnetic properties of the 1D chain. To conveniently describe the relative orientations of the physical quantities in the experiment by a Cartesian coordinate system, the plane of the Pt substrate is defined as the X-Y plane and the Co chain axis, along the X-axis. The easy axis, $K_u$, is then lying in the Y-Z plane with an inclination angle $\phi_K = 43°$ from the Z-axis. The external field, $H$, can be applied in any relative orientation with respect to the axis of anisotropy, $K_u$. Its orientation is specified by $(\theta,\phi)$ in which $\theta$ is the angular coordinate in the X-Z plane, and $\phi$, in the Y-Z plane. The Z-axis is then specified as $(\theta = 0, \phi = 0)$. Thus, the Hamiltonian can be written as

$$E_H = -\sum_{<i,j>} J\vec{S}_i \cdot \vec{S}_j - K_u \sum_i S_{iu}^2 - \mu \vec{H} \cdot \sum_i \vec{S}_i, \qquad (1)$$

where the first term represents the FM exchange interaction with the coupling constant $J > 0$, the index $<i,j>$ runs over all the pairs of nearest neighbour sites $i$ and $j$, and $S_i$ is the reduced spin variable at site $i$. The effective moment at each site for a Co atom is $\mu = 4\ \mu_B$. $K_u$ is the uniaxial anisotropy constant represents the on-site anisotropy for each Co atom. With $K_u > 0$, the axis of anisotropy defines the orientation of easy axis. $S_{iu}$ is the projective component of the reduced spin variable along the uniaxial direction, and, the applied magnetic field, $H$, is in units of Oe. In the calculation, each energy term, such as the anisotropy, magnetic energy, thermal energy, is expressed in units of the exchange coupling strength, $J$. It is 7 meV determined in the experiment [10]. This value is obtained by rough estimation that is also adopted in the previous simulation [12]. Although in a recent investigation by a finite-size transfer matrix approach, the value of $J$ is estimated differently as 20 meV [16], the result to describe the 1D monatomic chain of Co by the present simulation remains equally valid since it does not rely on the absolute value of $J$, which will be elaborated later.

To generate a trial spin configuration with one MC step, a spin is randomly chosen and then flipped towards an arbitrary orientation with a probability, $p = \text{Min}\{1, \exp(-\Delta E_H / k_B T)\}$, where the function, Min, is to pick the smallest value in the argument, $\Delta E_H$ is the change in energy owing to the reorientation of the spin at temperature, $T$. The computational time unit, $t_C$, for the simulation is then defined by the MC step per site (MCSS). For a chain of $N$ atoms, there are $N$ steps of MC calculation in one MCSS. For a data point in the hysteresis loop, the value of magnetization has been averaged over 200 independent runs to reduce the level of fluctuation.

## 3. Results and Discussions

*M-H* curves are calculated for the monatomic chain of Co with different parameters of $N$, $R$, and $K_u$, at various temperatures. The temperature dependent coercivity, $H_C(T)$, and the corresponding blocking temperature, $T_B^{Cal}$, are obtained corresponding to these various conditions. It is found that $H_C(T)$ follows a universal behaviour, $H_C(T)/H_C(T_{CL}) = h_0 (e^{T_B/T} - e)$ with $T_{CL} < T < T_B^{Cal}$. The experimentally observed results reported in Ref [10], *i.e.*, the temperature dependent coercivity, $H_C(T)$, and the angular dependence of the remanent magnetization, $M_R(\theta,\phi)$, are then reproduced by the present calculation method. The effect of the anisotropy and the atomic chain length on the blocking temperature is then assessed in detail.

*3.1. Temperature dependent coercivity $H_C(T)$ and blocking temperature*

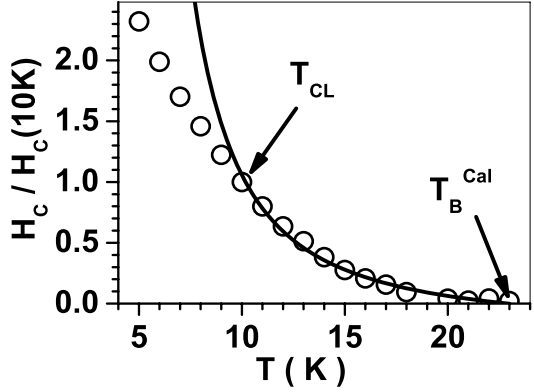

Figure 1 $H_C(T)$ determined from the M-H loops calculated with the conditions, $N = 80$, $R = 2.67 \times 10^{-2}$ Oe/MCSS, $K_u = 0.3$, and $J = 7$ meV. It is normalized to the value calculated at $T = 10$ K, i.e. $H_C(10\ K) = 17$ kOe.

For the atomic chain with experiment parameters of $N = 80$, $K_u = 0.3$, and $J = 7$ meV, the M-H curves at different temperatures have been calculated at the field sweeping rate of $R = 2.67 \times 10^{-2}$ Oe/MCSS. The temperature dependent coercivity $H_C(T)$ determined from these calculated loops is presented in figure 1. It is normalized to the coercivity at $T = 10$ K, $H_C(10K) \sim 17$ kOe. Taking into account the thermal activation effect, the coercivity is expected to decrease exponentially with increasing temperature, i.e., $H_C(T) \sim e^{T_B/T}$, and vanishes at the blocking temperature, $T_B = T_B^{Cal}$. The solid curve in figure 1 is a fitting result using the function

$$H_C(T)/H_C(T_{CL}) = h_0 (e^{T_B/T} - e), \qquad (2)$$

where $h_0$ is a constant and the exponential constant $e$ is obtained from the condition, $H_C(T_B^{Cal}) = 0$. The blocking temperature determined by the fitting is $T_B^{Cal} \sim 22.6$ K. It is higher than the experimental value, $\sim 15 \pm 5$ K, attributed to the faster field sweeping rate in the calculation.

Equation (2) reflects the thermal activation effect according to the Arrhenius law. It is expected as a direct consequence from the model adopted in the calculation. For the reorientation of a spin at temperature, $T$, the probability is modelled as $p = \mathrm{Min}\{1, \exp(-\Delta E_H / k_B T)\}$, with $\Delta E_H$, the change in energy owing to the spin flip. By treating $\Delta E_H$ as the energy barrier $E_a$, the time needed for the spin flip is, $t \propto 1/p \propto \exp(E_a/k_B T)$. Since $H_C(T) \propto t(T)$, and at $T = T_B$, the blocking effect is expected to vanish due to the activation of thermal energy, $k_B T_B \sim E_a$, we have $H_C(T) \propto (e^{T_B/T} - e)$. By accounting for the fact that the above property is valid at $T > T_{CL}$ due to the limitation of the calculation elaborated below, equation (2) is thus obtained. Equation (2) is not applicable in the temperature range $T < T_{CL} = 0.442 T_B^{Cal}$. For a system with a magnetic potential barrier $E_a$ which obstructs the magnetization reversal, the ratio, $E_a/k_B T$, determines the switching rate of the spins over the barrier subject to the thermal activation effect described by the Arrhenius law. However, Wernsdorfer *et al*, have shown clearly with an experiment that as the magnitude of $E_a/k_B T$ reaches 58.8, the quantum tunneling rate of the spins will become comparable to the thermal activation one for the 1D magnetic system in their experiment [20]. To avoid such a possible quantum effect which is beyond the description of the classical MC simulation, we applied the simulation only within the temperature region of $0.442 T_B^{Cal} < T < T_B^{Cal}$ in the following presented calculations.

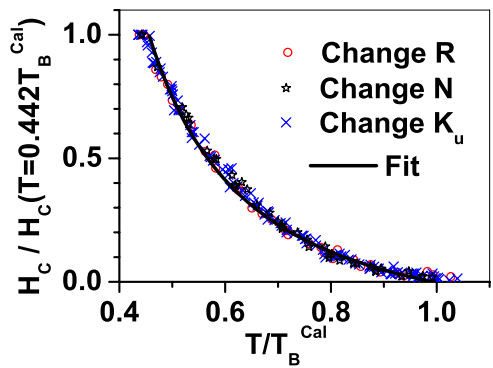

Figure 2 $H_C(T)$ normalized over the coercivity at $T = 0.442 T_B^{Cal}$, $H_C(0.442 T_B^{Cal})$, versus the reduced temperature, $T/T_B^{Cal}$ with one of the parameters, $N$, $V$, and $K_u$ in variation. The solid curve is for the fitting results by equation (2).

Since $T_B{}^{Cal}$ depends on $N$, $R$, and $K_u$, it is necessary to investigate its dependence on these parameters in order to understand the magnetization reversal behaviour or the property of $H_C(T)$. Figure 2 shows the temperature dependent coercivity in the temperature interval $0.442T_B{}^{Cal} < T < T_B{}^{Cal}$ obtained by a) increasing $R$ from $2\times10^{-3}$ Oe/MCSS to 1.34 Oe/MCSS with $N = 80$ and $K_u = 0.3J$, b) varying $N$ from 10 to 240 atoms with $K_u = 0.3J = 2.1$ meV and $R = 2.67\times10^{-2}$ Oe/MCSS, and c) varying $K_u$ from 0.1 to $1.32J$ by fixing $N = 80$ atoms and $R = 2.67\times10^{-2}$ Oe/MCSS. The coercivity thus calculated is then normalized over the value of $H_C$ at $T = 0.442T_B{}^{Cal}$ and plotted versus the reduced temperature $\tau = T/T_B{}^{Cal}$. As expected, these data reveal a universal behaviour. They collapse on a single curve. It indicates that the physical process manifested in the temperature variation effect for the spin chain is, indeed, the thermal activation against the anisotropy potential barrier. The fitting result of equation (2) is also plotted in figure 2 by the solid curve. In the temperature interval $0.442T_B{}^{Cal} < T < T_B{}^{Cal}$, equation (2) gives a good description on the result of our calculation. With this universal behaviour, the blocking temperature can be uniquely determined corresponding to specified $V$, $K_u$, and $N$ parameters.

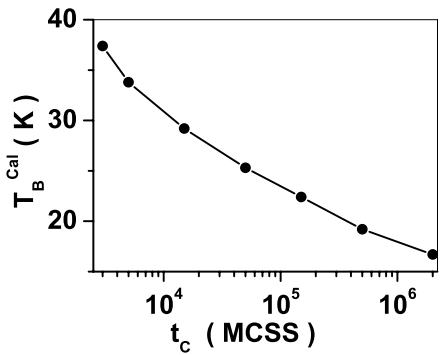

Figure 3 Blocking temperatures $T_B{}^{Cal}$ versus the computer calculation time, $t_C$. The solid curve is plotted to guide the eyes.

Figure 3 shows $T_B^{Cal}$ determined from the calculated *M-H* loop versus the computer calculation time, $t_C$, needed for sweeping a complete *M-H* loop, which is in units of MCSS. The solid curve connecting the "data" points in figure 3 is a guide for the eyes. For a slow relaxation process, it is expected that the system relaxes exponentially with the physical time, *i.e.* it shows a linear behaviour in the log(*t*) plot. Although $t_C$ is not linearly proportional to the physical time [21-23], the calculated blocking temperature $T_B^{Cal}$ versus $t_C$ still exhibits a behaviour close to an exponential relaxation process, *i.e.*, close to a linear relation in the log($t_C$) plot, as shown by figure 3. This apparent higher order effect deviating from the logarithmically exponential relaxation with the computer timescale is probably arising from the nonlinear conversion relation from the computer calculation time to the physical timescales. Without affecting the conclusion in the present work, the computer calculation timescale is adopted. By increasing the field sweeping time duration for a complete *M-H* loop from $3\times10^3$ MCSS to $2\times10^6$ MCSS, corresponding to reducing the field sweeping rate from 1.34 Oe/MCSS to $2\times10^{-3}$ Oe/MCSS, the blocking temperature then decreases from 37.4 K to 16.7 K. The dependence of the coercivity, hence, the blocking temperature $T_B^{Cal}$, upon the field sweeping rate, *R*, indicates that the calculated FM property at $T < T_B^{Cal}$ is not for an equilibrium state. The occurrence of a finite-area hysteresis, which is calculated based on the model of 1D atomic chain with SREI, does not conflict with the Mermin-Wagner theorem since the calculated hysteresis loop is not for a true FM property with a long range order in thermodynamic equilibrium.

*3.2. Angular dependent hysteresis loops and remanent magnetization, $M_R(\theta,\phi)$*

Experimentally, the coercivity and the blocking temperature of the 1D system also depend on the measured time or the field sweeping rate [20]. For a comparison with the results reported in the experiment, *M-H* curves at $T$ = 10 and 45 K are calculated in the present work with $N$ = 80 atoms, $K_u$ = 0.3$J$ = 2.1 meV, and $R$ = 2.67×10$^{-2}$ Oe/MCSS, as shown in figure 4. The magnetization *M* is expressed in units of 4$\mu_B$/atom in both figures 4a and 4b while the applied field *H*, in units of $H_{max}$, which is 120 kOe in figure 4a and 600 kOe in figure 4b. The calculations have reproduced the major features of the *M-H* loops observed in the experiment, *i.e.*, an open hysteresis loop at $T$ = 10 K and a SPM behaviour at $T$ = 45 K. In addition, the remanent magnetizations $M_R(\theta,\phi)$ determined from the calculated loops at $T$ = 10 K exhibit the same angular dependence on $\theta$ and $\phi$ as observed in the experiment. All the calculated magnetizations are expressed in units of 4 $\mu_B$/atom.

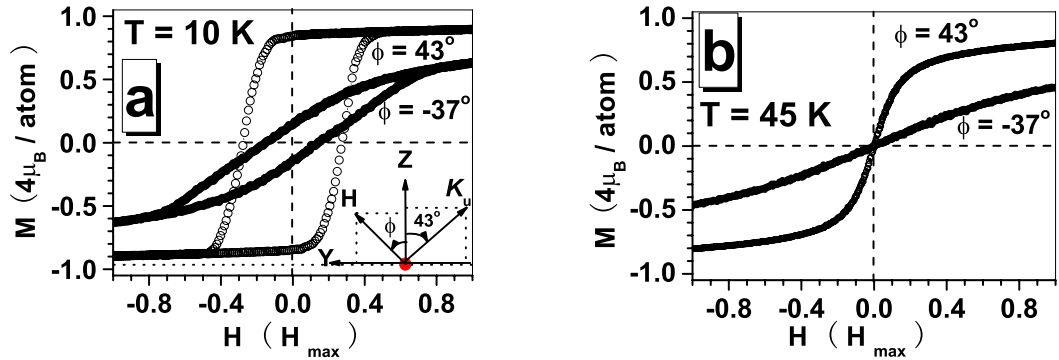

Fig. 4 *M-H* curves calculated for the 1D monatomic chain with $N$ = 80 atoms. The magnetization *M* is expressed in units of 4$\mu_B$/atom. The applied field is expressed in units of the maximum sweeping limit, $H_{max}$, in the calculation. (a) *M-H* curves at $T$ = 10 K. $H_{max}$ = 120 kOe. The lower right quarter shows a simple depiction for the relative orientations of the monatomic chain axis, the anisotropy axis, the applied field direction and the coordinate axes. (b) The curves calculated at $T$ = 45 K. $H_{max}$ = 600 kOe.

To clearly reveal the details in calculating the *M-H* curves for the monatomic chain,

the relevant axes are depicted in a simple diagram shown in the lower right quarter of figure 4a. By the depiction, the Y-Z plane is lying in the plane of the figure with the positive Y directed to the left and the positive Z upwards. The axis of the 1D chain is along the X-axis whose positive direction points into the plane of the figure. According to the experiment, the easy axis of anisotropy, $K_u$, lies in the Y-Z plane with an inclination, $\phi_K = 43°$, from the Z-axis. The *M-H* curves shown in figure 4a for $T = 10$ K and figure 4b for $T = 45$ K are calculated using the experimental conditions, *i.e.*, the field is applied along the easy axis with $\phi = \phi_K = 43°$, and at 80° away from the easy axis with $\phi = -37°$. At $T = 10$ K, both the two calculated curves exhibit open hysteresis loops. As the temperature rises to $T = 45$ K, the *M-H* curves reveal the property of SPM, see figure 4b. These reproduce the experimental features [10].

A further investigation by calculation on the remanent magnetization, $M_R(\theta,\phi)$, is performed at $T = 10$ K. In the experiment, $M_R(\theta,\phi)$ is the magnetization measured by a small field after the magnetizing field, which drives the sample to saturation, is removed [10]. It is actually the remanent magnetization in a *M-H* measurement. Its magnitude depends on the orientation $(\theta,\phi)$ of the sweeping applied field. In figure 5a, $\phi$ is defined with the same meaning as in figure 4a and $\theta$ is defined as the inclination angle from the Z-axis in the X-Z plane as shown by the depiction in figure 5b. The remanent magnetization, $M_R(\theta,\phi)$, versus the inclination angle, $\phi$, in the Y-Z plane is determined from a series of calculated *M-H* curves and plotted in figure 5a by the solid circles. Similarly, the calculated $M_R$ versus $\theta$ is presented in figure 5b. The solid curve in figures 5a and 5b are for the functions $g(\phi) = |\cos(\phi-\phi_K)|$ and $f(\theta) = |\cos(\theta)|$, respectively, which are determined in the experiment by fitting the measured remanent magnetization. The calculated $M_R$ at $T = 10$ K, as shown in figure 5, describes well the angular dependent property of $M_R$ measured in the experiment [10]. Interestingly, even at temperature as low as 10 K, $H_C$ and $M_R$ are zero and the chain shows a SPM behaviour with the applied field $H$ exactly perpendicular to the easy axis, $K_u$, according to the present calculation. This is mainly due to the absence of an anisotropy component within the plane of the sweeping field, *i.e.* without an anisotropy barrier to obstruct the magnetization reversal process. A similar property is also observed with a FM material with large anisotropy, for example, for a FM nanowire with large shape anisotropy along the axis of the wire, $H_C$ and $M_R$ are also much reduced with the applied field $H$ perpendicular to the easy axis [24].

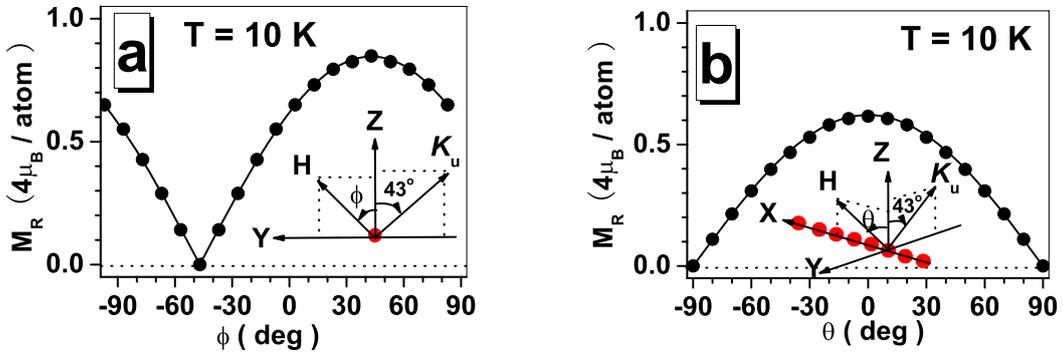

Figure 5 Remanent magnetization, $M_R(\theta,\phi)$, versus the orientation of the applied field at $T = 10$ K. $M_R$ is expressed in units of $4\mu_B$/atom. (a) $M_R$ versus $\phi$, with $\theta = 0°$. (b) $M_R$ versus $\theta$ with $\phi = 0°$. The solid curves in (a) and (b) are the fitting results of $|\cos(\phi-43°)|$ and $|\cos(\theta)|$, respectively.

*3.3. The dependence of blocking temperature on anisotropy constant and chain length*

Figure 6 shows $T_B^{Cal}$ versus the magnitude of the anisotropy constant, $K_u$ at a field sweeping rate of $R = 2.67\times10^{-2}$ Oe/MCSS. The solid circle is for $K_u = 2.1$ meV $= 0.3J$, which is the same as that in the experiment [10], and the solid triangle, $K_u$ = 9.3 meV = 1.32$J$, which is for the upper limit of anisotropy with a single Co atom on Pt (997) [14]. According to the analysis by Li *et al* [12], the potential barrier in the anisotropic Heisenberg model can be approximated by the expression, $E_a = [2K_u + (\sum_j JS_j + \mu H)S_i]^2 / 4K_u$. This energy barrier depends not only on the anisotropy constant $K_u$ but also on the exchange coupling strength $J$ and the chain length, $N$. The barrier height increases with the increasing anisotropy when $K_u > \frac{1}{4}(\sum_j JS_j + \mu H)S_i$. In the limit $K_u \gg \frac{1}{4}(\sum_j JS_j + \mu H)S_i$, $E_a$ is expected to increase linearly with the magnitude of $K_u$. This is consistent with our result by a direct MC calculation. The increase of $T_B^{Cal}$ with $K_u$ is linear as $K_u > 0.3J$.

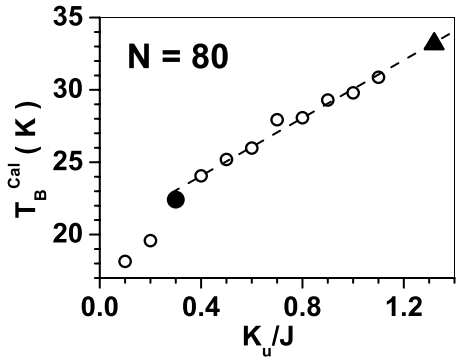

Figure 6 Blocking temperatures $T_B^{Cal}$ versus the anisotropy constant, $K_u$. The field sweeping rate is $R = 2.67 \times 10^{-2}$ Oe/MCSS.

Figure 7 shows the dependence of $T_B^{Cal}$ on $N$ for the monatomic chain with $K_u = 0.3J$ and $R = 2.67 \times 10^{-2}$ Oe/MCSS. It reveals that $T_B^{Cal}$ increases with $N$, and then becomes almost a constant after reaching 80 atoms. In a previous numerical investigation on a classical spin chain with similar Hamiltonian to our model, three magnetization reversal modes have been proposed for the 1D system [25]. As the length of the spin chain increases from several spins to hundreds of spins, the reversal modes change from coherent rotation in which all the spins in the chain rotate coherently, to the soliton-antisoliton nucleation mode with which the magnetization reversal proceeds with the spins splitting into two parts with opposite direction of magnetization reversal, and finally to the multidroplet nucleation mode with which many nuclei appear at the same time and then join each other, leading to a complete magnetization reversal. In our simulations, we observe the reversal process by coherent rotation in the small size limit of the monatomic chain. By increasing the chain length, the magnetization reversal of the chain proceeds more often by the soliton-antisoliton nucleation mode in a series of repeated calculations using the same condition. As the chain length further increases exceeding the critical value of 80 atoms, the reversal is always by multidroplet nucleation mode which is similar to the reversal mode of an infinite chain. Thus, $T_B$ becomes almost independent of the number of atoms with $N > 80$, as shown in figure 7. According to the present calculation, before $N$ approaches the critical value for the multidroplet nucleation mode, the reversal process shows a property of bistability either with the coherent rotation or the soliton-antisoliton nucleation, and $T_B$ is obtained as a result of statistical average.

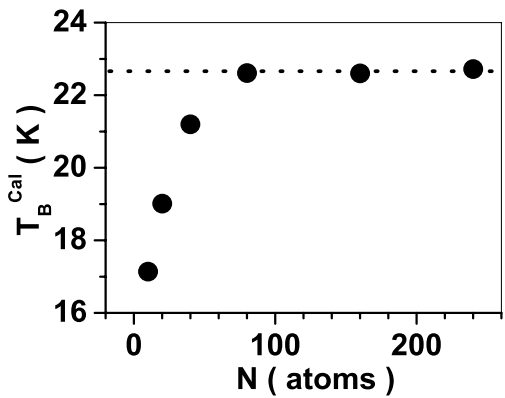

Figure 7 Dependence of $T_B^{Cal}$ on the number of atoms, $N$, calculated with $K_u = 0.3J$. The field sweeping rate is $R = 2.67 \times 10^{-2}$ Oe/MCSS.

## 4. Conclusion

In conclusion, we have calculated the magnetic properties of a 1D monatomic chain of Co by a classical MC calculation based on the anisotropic Heisenberg model. The temperature dependence of the coercivity, $H_C(T)$, and the angular dependence of the remanent magnetization, $M_R(\theta,\phi)$, upon the applied field direction are in agreement with the previously reported experiment. The calculation is demonstrated as a proper technique to calculate the properties of a 1D spin chain with a slow relaxation process. It is apparent that the potential barrier obstructing the spin reversal is a crucial factor, resulting in an FM like *M-H* behaviour for the 1D monatomic chain at low temperature. Additionally, the calculation result does not conflict with the traditional 1D spin lattice theory with SREI [1,4,5]. This can be well described by the statement put forward by Jacobs and Bean [26] '' . . . *the one-dimensional Ising chain* (*or any anisotropic chain*) *is not ferromagnetic in equilibrium but will, in fact, show all the usual characteristics of ferromagnetic matter owing to the difficulty of reaching equilibrium.*''

References

e-mail: cpchen@pku.edu.cn, phone : +86-10-62751751